\documentclass[showpacs,twocolumn,prl,amsmath,amssymb]{revtex4}

\usepackage{graphicx}
\usepackage{dcolumn}
\usepackage{bm}
\begin{document}
\title{
Coexistence of double-{\bf Q} spin density wave and
multi-{\bf Q} pair density wave\\
in cuprate oxide superconductors
}
\author{Fusayoshi J. Ohkawa}
\affiliation{Division of Physics, Graduate School of
Science,  Hokkaido University, Sapporo 060-0810, Japan}
\email{fohkawa@phys.sci.hokudai.ac.jp}
\date{\today}
%
\begin{abstract} 
Spatial $4a\times4a$ modulations,
with $a$ the lattice constant of CuO$_2$ planes,
or the so called checkerboards 
can arise from  double-{\bf Q} 
spin density wave  (SDW) with ${\bf Q}_1 = (\pm\pi/a, \pm3\pi/4a)$ and 
${\bf Q}_2 = (\pm3\pi/4a, \pm\pi/a)$.  
When multi-{\bf Q} pair density wave, that is,  
the condensation of $d\gamma$-wave Cooper pairs with
zero total momenta, $\pm2{\bf Q}_1$, $\pm2{\bf Q}_2$,
$\pm4{\bf Q}_1$, $\pm4{\bf Q}_2$, and so on
is induced by the SDW, 
gaps can have fine structures similar to
those of the so called zero-temperature pseudogaps.
\end{abstract}
\maketitle

High critical-temperature (high-$T_c$) superconductivity in cuprate oxides
is an important  and long standing issue  since its discovery \cite{bednorz}.
Many unconventional properties are observed in addition to high $T_c$:
the so called spin gap \cite{spingap} or pseudogap above $T_c$
\cite{Ding,Shen2,Shen3,Ino,Renner,Ido1,Ido2,Ekino},
$4a \times 4a$ spatial modulations called checkerboards,
with $a$  the lattice constant of CuO$_2$ planes,  of
local density of states (LDOS)   
around vortex cores \cite{vortex}, 
above $T_c$  \cite{Vershinin},  and
below $T_c$ \cite{Hanaguri,Howald,Momono}, 
the so called 
zero-temperature pseudogaps (ZTPG)  \cite{McElroy}, and so on.
%
The issue  cannot be settled unless 
not only high $T_c$ 
but also  unconventional  properties are explained within a
theoretical framework.

Since cuprates are highly anisotropic,
thermal critical superconducting (SC) fluctuations,
which would be divergent at nonzero $T_c$ in  two dimensions
\cite{Mermin},
must play a role in the opening of pseudogaps.
This issue is examined in another paper \cite{PS-gap}. 

The period of  checkerboard modulations  is independent of energies.
Their amplitude  depends on energies;
it is only large in the gap region. 
When the modulating part is divided into symmetric and
asymmetric ones with respect to the chemical potential, 
the symmetric one is larger than the asymmetric one
\cite{Howald}. Fine structures are observed in ZTPG  \cite{McElroy}.
It is difficult to explain these observations 
in terms of  charge density wave (CDW).
Several possible  mechanisms have been proposed:
Fermi surface nesting \cite{Davis},
valence-bond solids \cite{Sachdev,Vojta},
pair density waves \cite{halperin,cdw}, 
hole-pair \cite{Davis,Chen1,Chen2} 
or single-hole \cite{Davis,Sachdev} Wigner solids,
and  Wigner supersolids \cite{Tesanovic,Anderson}.
The purpose of this Letter is to propose another mechanism:
The spatial modulation of LDOS is due to
spin density wave (SDW) and ZTPG is due to
the coexistence of SDW and superconductivity or
pair density waves induced by SDW.

Cuprates with no dopings are Mott-Hubbard insulators, which exhibit
antiferromagnetism.  As dopings are increased, 
they exhibit the Mott-Hubbard  transition or crossover and they becomes metals.
High-$T_c$ superconductivity occurs in such a metallic phase.
According to a previous theory \cite{slave}, which is consistent with 
a combined one of Hubbard's \cite{Hubbard} 
and Gutzwiller's \cite{Gutzwiller} ones,
a three-peak structure appears in the density of states;
the so called Gutzwiller's quasiparticle band  between
the lower and upper Hubbard bands (LHB and UHB). 
This is confirmed by the single-site approximation (SSA) 
\cite{Mapping-1,Mapping-2} that
includes all the single-site terms or by the dynamical mean-filed theory (DMFT)
\cite{georges}. 
The three-peak structure corresponds to the so called
Kondo peak between two subpeaks
in the Anderson model or in the Kondo problem.
A Kondo-lattice theory is useful to treat strong correlations
 in the vicinity of the Mott-Hubbard transition \cite{disorder}.
The superexchange interaction, which arises from the virtual exchange
of pair excitations of electrons across LHB and UHB
even in the metallic phase,
favors an antiferromagnetic (AF) ordering
of $(\pm\pi/a, \pm\pi/a)$.
Another exchange interaction arises from that 
of Gutzwiller's quasiparticles.
Since  Fermi surface nesting
plays a role,  this exchange interaction favors AF
orderings of  nesting wave numbers.
Then, one of the most plausible scenarios for the modulating LDOS
is that double-{\bf Q} SDW with
${\bf Q}=
(\pm\pi/a, \pm3\pi/4a)$ and $(\pm3\pi/4a, \pm\pi/a)$
are stabilized and 
the $2{\bf Q}$ modulation of LDOS is one of their second-harmonic effects;
$2{\bf Q}=(\pm 3\pi/2a, \pm2\pi/a)$
 or  $(\pm 2\pi/a, \pm3\pi/2a)$ is equivalent to
 $(\pm \pi/2a, 0)$ or  $(0, \pm\pi/2a)$. 
On the other hand,
a  magnetic exchange interaction can be a Cooper-pair interaction \cite{hirsch}.
According to previous papers \cite{KL-theory1,KL-theory2}, 
the superexchange interaction as large as $|J|=0.1\mbox{-}0.15$~eV
is strong enough to reproduce observed SC $T_c$;
theoretical  $T_c$ of  $d\gamma$ wave are definitely much higher 
than those of other waves. 
Since the two intersite exchange interactions are
responsible for not only antiferromagnetism
but also  superconductivity, 
the coexistence of antiferromagnetism and superconductivity 
or the competition between them must be
a key for explaining unconventional properties of cuprate oxide
superconductors.

In order to demonstrate
 the essence of the mechanism,
we consider  first a  mean-field Hamiltonian on the square lattice,
that is,  non-interacting electrons 
in the presence of AF and $d\gamma$-wave SC fields:
%
${\cal H} =  \sum_{{\bf k}\sigma} E({\bf k})a_{{\bf k}\sigma}^\dag 
a_{{\bf k}\sigma} + {\cal H}_{AF} +{\cal H}_{SC}$.
%
The first term describes non-interacting electrons;
%
$E({\bf k}) = - 2t \left[\cos(k_xa) + \cos(k_ya) \right]
 -4 t^\prime \cos(k_xa) \cos(k_ya) - \mu $,
 %
 with $t$ and $t^\prime$ transfer integrals between nearest and next-nearest
 neighbors and $\mu$ the chemical potential, 
 is the dispersion relation of electrons.
 We assume that $t^\prime/t = -0.3$.
 The second term describes AF fields with
 wave number  ${\bf Q}=(\pm 3\pi/4a, \pm\pi/a)$
 or  $(\pm \pi/a, \pm3\pi/4a)$:
\begin{eqnarray}
{\cal H}_{AF} &=& 
- \sum_{{\bf k}\sigma\sigma^\prime}\sum_{\xi=x,y,z}
\sigma_{\xi}^{\sigma\sigma^\prime}  \left( 
\Delta_{\xi} a_{{\bf k}+{\bf Q}\sigma}^\dag a_{{\bf k}\sigma^\prime}
\right.
\nonumber \\ && \hspace*{2cm} \left.
+ \Delta_{\xi}^{*} a_{{\bf k}-{\bf Q}\sigma}^\dag a_{{\bf k}\sigma^\prime}
\right) ,
\end{eqnarray}
with $\sigma_{\xi}^{\sigma\sigma^\prime}$ the Pauli matrixes.
A single-${\bf Q}$ 
structure  or the so called stripe is assumed 
for the sake of simplicity.
The origin of real space is chosen in such a way that
$\Delta_{\xi}$ and $\Delta_{\xi}^{*}$ are real and positive;
the external filed is 
$\Delta_{\xi}({\bf R}_{i}) = 2 \Delta_{\xi} \cos({\bf Q}\cdot{\bf R}_i) $.
%
The Brillouin zone is folded by periodic AF fields.
When we take its origin 
at a  zone boundary of a folded zone, 
electron pairs of ${\bf k}+l {\bf Q}$ and
$-{\bf k}+l {\bf Q}$ can be bound  \cite{com}, 
with $l $ being an integer.
We assume the following $d\gamma$-wave SC fields:
\begin{eqnarray}
{\cal H}_{SC}   &=&
- \frac1{2} \sum_{\bf k} \eta_{d\gamma}({\bf k}) 
\sum_{l} \Bigl(
\Delta_{l} a_{{\bf k}+l{\bf Q} \uparrow}^\dag 
a_{-{\bf k}+l{\bf Q}\downarrow}^\dag 
\nonumber \\  && \hspace*{2cm}
+ \Delta_{l}^{*} a_{-{\bf k}+l{\bf Q}\downarrow} 
a_{{\bf k}+l {\bf Q}\uparrow} 
\Bigr) ,
\end{eqnarray}
with
$\eta_{d\gamma} ({\bf k}) = \cos(k_xa) - \cos(k_ya)$. 
%
The global phase of single-particle wave functions can be chosen
in such a way that $\Delta_{0}$ and $\Delta_{0}^{*}$ are real and positive.
We assume 
%
$\Delta_{l} = \Delta_{-l}$
%
for $l \ne 0$  for the sake of simplicity,
although we have no argument that other cases can be excluded.

The homogeneous part of LDOS per spin is given by
\begin{equation}
\rho_{0} (\varepsilon)
= - \frac1{2\pi N} \sum_{{\bf k}\sigma} \mbox{Im}\left[
G_{\sigma}(\varepsilon + i\gamma, {\bf k}; 2 l{\bf Q}) 
\right]_{l=0} ,
\end{equation}
with $\gamma/|t| \rightarrow +0$, 
where $G_{\sigma} (\varepsilon+i\gamma, {\bf k};2l{\bf Q}) $
is the analytical continuation from the upper half plane of
\begin{equation}
G_{\sigma} (i\varepsilon_n, {\bf k}; 2l{\bf Q}) =
- \int_{0}^{\beta} \hspace{-6pt}
d\tau e^{i \varepsilon_n \tau} \left<T_{\tau}
a_{{\bf k}-l{\bf Q}\sigma}(\tau) a_{{\bf k}+l{\bf Q} \sigma}^\dag 
\right> ,
\end{equation}
with $\beta = 1/k_BT $; we assume $T=0$~K so that $\beta\rightarrow +\infty$.
The modulating part with wave number $2l{\bf Q}$ is given by
\begin{eqnarray}\label{rho1A}
\rho_{2l{\bf Q}}(\varepsilon;{\bf R}_{i}) \hspace{-2pt} &=&
\hspace{-2pt} - \frac1{2\pi N} \hspace{-2pt}
\sum_{{\bf k}\sigma} \mbox{Im} \! \left[
e^{i2l{\bf Q}\cdot{\bf R}_i}
G_{\sigma} (\varepsilon \!+\! i\gamma, {\bf k};2l{\bf Q}) 
\right.
\nonumber \\ &&   \left.
+ e^{-i2l{\bf Q}\cdot{\bf R}_i}
G_{\sigma} (\varepsilon+i\gamma, {\bf k};-2l{\bf Q}) 
\right] ,
\end{eqnarray}
Since $\Delta_{\xi}$ and $\Delta_{0}$ are real
and  $\Delta_{l} = \Delta_{-l}$ for $l \ne 0$, 
$G_{\sigma} (\varepsilon+i\gamma, {\bf k};2l{\bf Q}) =
G_{\sigma} (\varepsilon+i\gamma, {\bf k};-2l{\bf Q})$.
Then, 
Eq.~(\ref{rho1A}) becomes simple in such a way that
\begin{equation}
\rho_{2l{\bf Q}}(\varepsilon;{\bf R}_{i})=
2\cos(2 l {\bf Q}\cdot{\bf R}_{i}) \rho_{2l {\bf Q}}(\varepsilon) ,
\end{equation}
with
\begin{equation}\label{Eq2Dim}
\rho_{2l{\bf Q}} (\varepsilon)
= - \frac1{2\pi N} \sum_{{\bf k}\sigma} \mbox{Im}
G_{\sigma} (\varepsilon + i\gamma, {\bf k};\pm2l{\bf Q}) .
\end{equation}
The modulating part with $(2l+1){\bf Q}$ vanishes because
the up and down spin components cancel each other.
%

\begin{figure*}
\centerline{
\includegraphics[width=8.0cm]{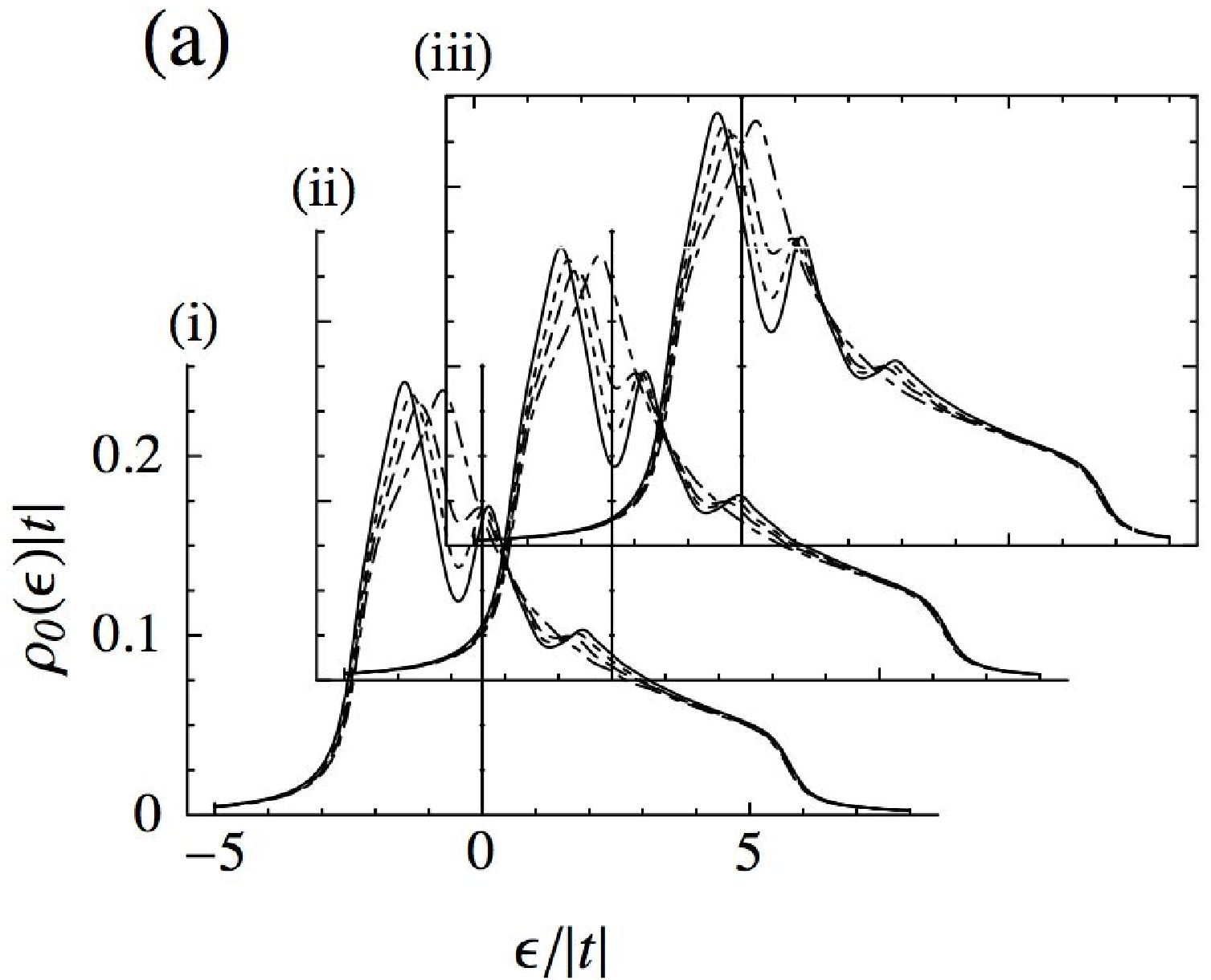}
\includegraphics[width=8.0cm]{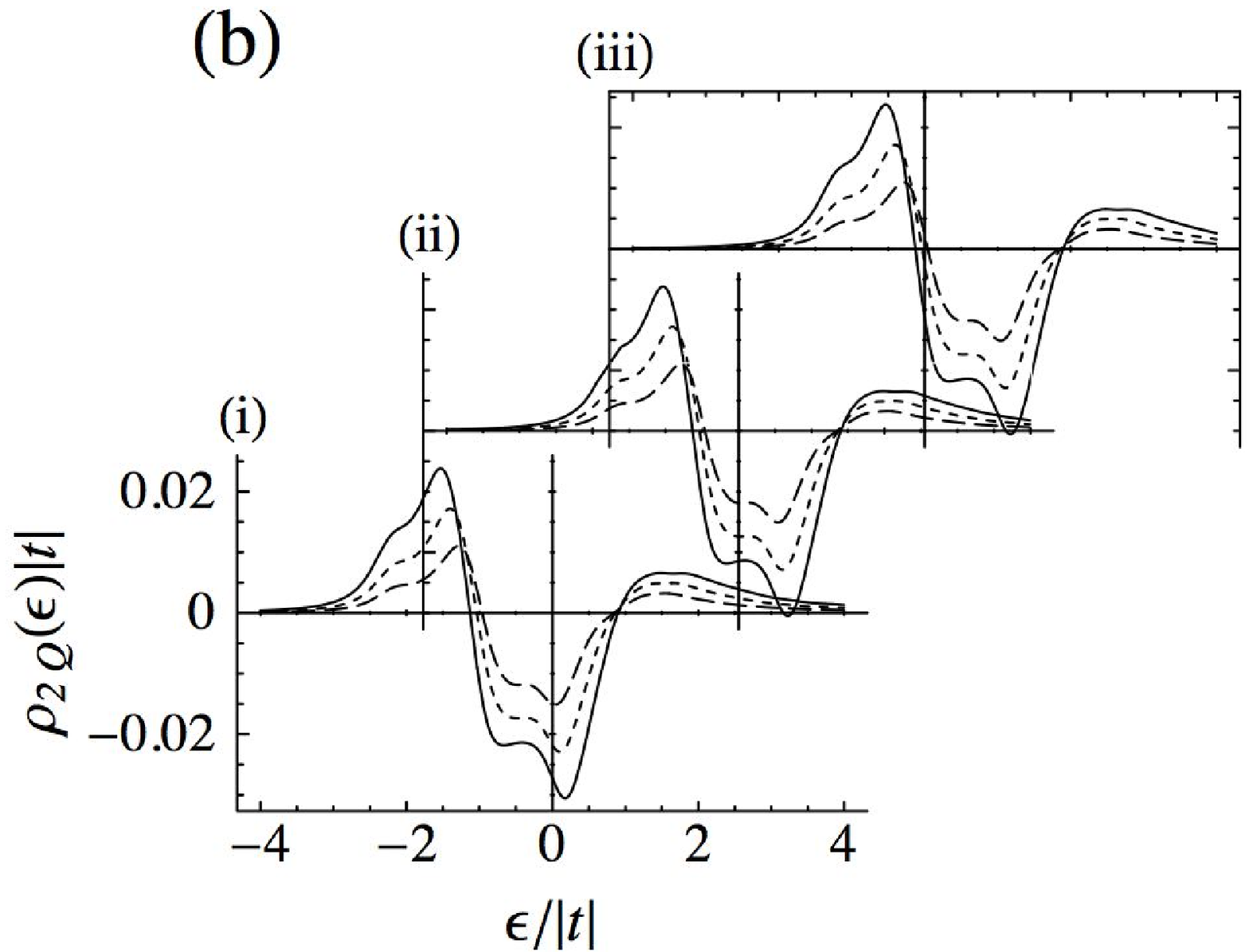}
}
\caption[1]{
(a) $\rho_{0}(\varepsilon)$ and (b) $\rho_{2{\bf Q}}(\varepsilon)$
in the presence of AF fields along the $x$ axis and no SC fields.
(i) $\mu/|t|=-0.5$, (ii) $\mu/|t|=-1$, and (iii) $\mu/|t|=-1.5$.
Solid, dotted, dashed, and chain lines are for $\Delta_x/|t| =0.5$, 0.4, 0.3, and 0,
respectively.
}
\label{sdw}
\end{figure*}
\begin{figure}
\centerline{
\includegraphics[width=8.0cm]{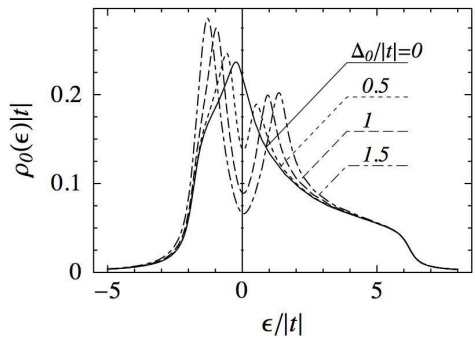}
}
\caption[2]{
$\rho_{0}(\varepsilon)$ 
in the presence of SC fields ($\Delta_0 \ne 0$ and $\Delta_l =0$ for $l\ne 0$)
and no AF fields;
$\mu/|t|=-1$ is assumed
for the chemical potential.
Since we assume nonzero $\gamma/|t|=0.3$,
$\rho_0 (\varepsilon=0)$ is nonzero.
}
\label{sc}
\end{figure}
\begin{figure*}
\centerline{
\includegraphics[width=8.5cm]{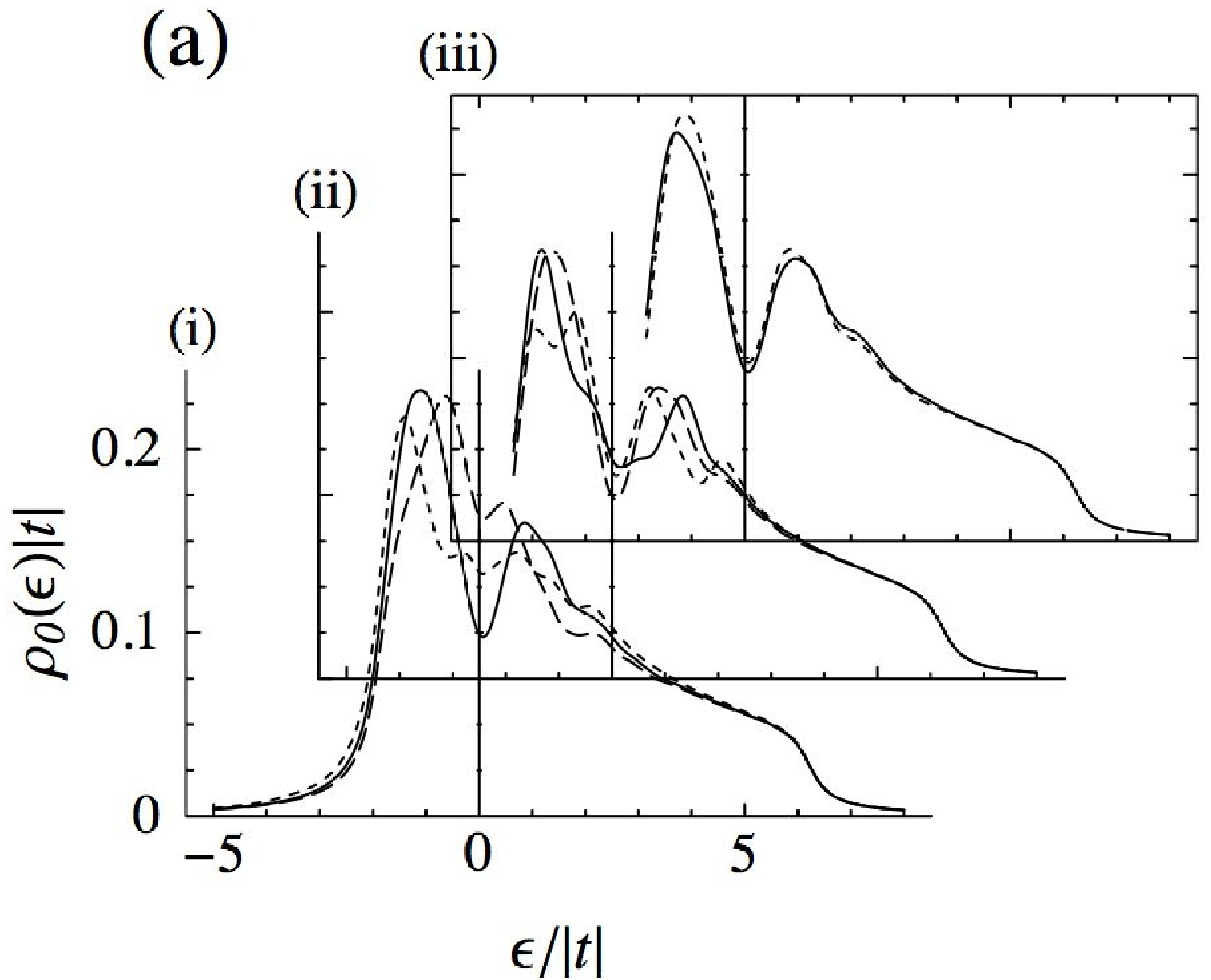}
\includegraphics[width=8.5cm]{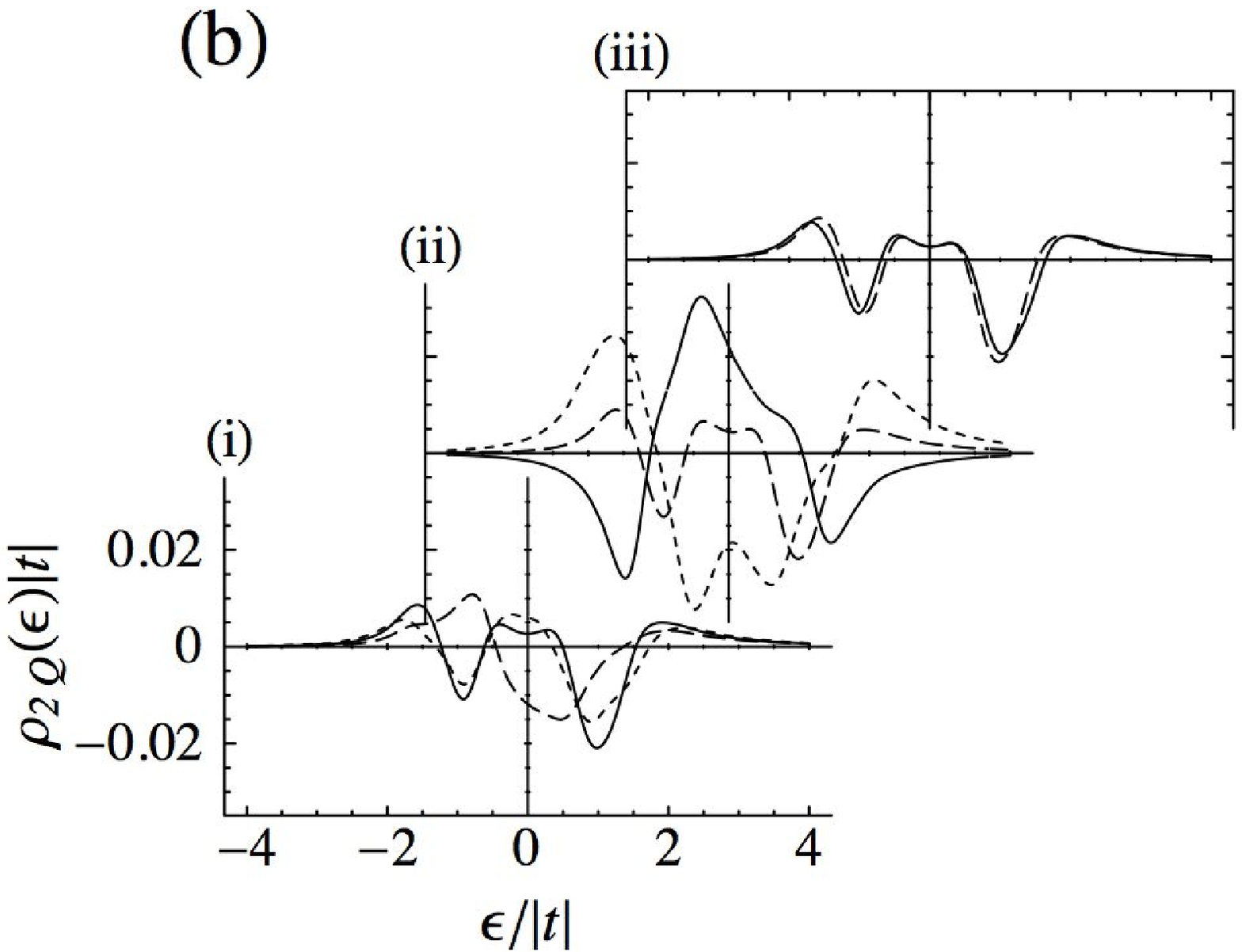}
}
\caption[3]{
(a) $\rho_{0}(\varepsilon)$ and (b) $\rho_{2{\bf Q}}(\varepsilon)$
 in the presence of AF fields 
of $\Delta_x /|t|=0.3$
and various SC fields; $\mu/|t|=-1$ is assumed
for the chemical potential.
In  (i),
solid, dotted, and dashed lines are for 
$(\Delta_0,\Delta_1,\Delta_2)/|t|= (1,0,0)$, 
$(0,1,0)$, and $(0,0,1)$.
In (ii), they are for
$(\Delta_0,\Delta_1,\Delta_2)/|t|= (1,0.5,0)$, $(1,-0.5,0)$, and $(1,0,\pm0.5)$,
respectively;
the result for $(1,0,-0.5)$ is the same as that for  $(1,0,0.5)$
within numerical accuracy.
In  (iii), solid and dotted lines are for
$(\Delta_0,\Delta_1,\Delta_2)/|t|=(1,\pm 0.5i,0)$ 
and $(1,0,\pm 0.5i)$, respectively;
the result for $(1,-0.5i,0)$ is the same as that for  $(1,0.5i,0)$
within numerical accuracy, and so on.
}
\label{sdw-sc}
\end{figure*}

We assume that $\Delta_{l} = 0$ for $| l|\ge 3$ and
we take a $5 \!\times\!2\!\times\!2$-wave approximation;
we consider couplings among single-particle excitations such as
electronic ones of 
$a^\dag_{{\bf k},\pm\sigma} \left|0\right>$, 
$a^\dag_{{\bf k}\pm{\bf Q,}\pm\sigma} \left|0\right>$,
and $a^\dag_{{\bf k}\pm2{\bf Q},\pm\sigma} \left|0\right>$
and hole-like ones of
$a_{-{\bf k},\pm\sigma} \left|0\right>$, 
$a_{-{\bf k}\pm{\bf Q,}\pm\sigma} \left|0\right>$,
and $a_{-{\bf k}\pm2{\bf Q},\pm\sigma} \left|0\right>$,
with
$\left|0\right>$ being the Fermi vacuum. 
Matrixes to be diagonalized are $20\times20$.
The  transformation diagonalizing them
is nothing but a generalized Bogoliubov transformation.
For the sake of numerical processes, nonzero $\gamma$ is
assumed: We assume  $\gamma/|t|=0.3$
instead of $\gamma/|t|\rightarrow +0$.

Figure~\ref{sdw}  show $\rho_0 (\varepsilon)$ and
$\rho_{2{\bf Q}} (\varepsilon)$ in the presence of
AF fields along the $x$ axis and no SC  fields;
results do not depend on the direction of AF fields.
A gap minimum is close to the chemical potential for
$\mu/|t|=-1$. 
This implies that the nesting wave number must be very close
to ${\bf Q}$ for $\mu/|t|=-1$. 
We assume $\mu/|t|=-1$ in the following part.
No fine structure can be seen in 
the low-energy part of $\rho_0 (\varepsilon)$.
The symmetric part of $\rho_{2{\bf Q}} (\varepsilon)$
is much larger than the asymmetric one.
The amplitude of CDW is small,
because positive and negative parts below the chemical potential
$(\varepsilon<0)$ cancel
largely each other.
 
Figure~\ref{sc}  show $\rho_0 (\varepsilon)$ in the presence of
SC fields and no AF fields. In the absence of AF fields,
even if $\Delta_{l} \ne 0$ for $l \ne 0$,
there is no modulating part in LDOS.

Figure~\ref{sdw-sc}  show $\rho_0 (\varepsilon)$ and
$\rho_{2{\bf Q}} (\varepsilon)$ in the presence of
AF fields along the $x$ axis and SC fields;
results do not depend on the direction of AF fields either.
Gaps can have fine structures.
The modulating part $\rho_{2{\bf Q}}(\varepsilon)$ can be quite
different 
between Figs.~\ref{sdw} and \ref{sdw-sc}
or in the  the absence and presence of SC fields.

In order to explain   checkerboards  and ZTPG 
in cuprates,
various extensions have to be made.
First of all, strong electron correlations 
in the vicinity of the Mott-Hubbard transition 
should be seriously considered; we had better use
the so called $d$-$p$ model or the $t$-$J$ model 
\cite{ZhangRice}.
A Kondo-lattice theory can treat such strong electron correlations
 \cite{disorder}.
Non-interacting electrons in this Letter correspond to
Gutzwiller's quasiparticles;  
observed specific heat coefficients  as large as 
$14$~mJ/K$^2$mol \cite{gamma1} imply
$|t|\simeq 0.04$~eV.
External AF and SC fields in this Letter correspond to 
conventional Weiss's static mean fields due to 
the superexchange interaction and the exchange interaction arising from
the virtual exchange of pair excitations of Gutzwiller's quasiparticles.

In  the orthorhombic or square lattice,
${\bf Q}_1 = (\pm\pi/a$, $\pm3\pi/4a)$ and 
${\bf Q}_2 = (\pm3\pi/4a, \pm\pi/a)$ are equivalent to each other.
Since the Fermi surface nesting is  sharp, 
double-{\bf Q} SDW must be
 stabilized in cuprates rather than single-{\bf Q} SDW;
magnetizations  of the two waves 
must be orthogonal to each other \cite{multi-Q}.
We propose that checkerboards in the absence of SC order parameters
are due to the second-harmonic effect of double-{\bf Q} SDW.

In the presence of double-{\bf Q} SDW, 
 superconductivity should  be extended to include 
Cooper pairs 
whose total momenta are zero, $\pm2{\bf Q}_1$, $\pm2{\bf Q}_2$,
$\pm4{\bf Q}_1$, $\pm4{\bf Q}_2$, and so on. 
Not only checkerboards but also  ZTPG 
can  arise in the coexistence phase of double-{\bf Q} SDW
and multi-{\bf Q}  superconductivity.
The solid line in Fig.~\ref{sdw-sc}(a)-(ii)
resembles to observed fine structures  of ZTPG,
though single-{\bf Q} SDW  is assumed there.
The  observed ZTPG  phase may be characterized by
a precise comparison of $\rho_0 (\varepsilon)$ and
$\rho_{2{\bf Q}} (\varepsilon)$ between observations and theoretical results 
for various parameters for double-{\bf Q} SDW and multi-{\bf Q}
superconductivity.

Although AF fluctuations are well developed in the  checkerboard 
and the ZTPG phases,
it is not certain that AF moments are actually present there. 
Checkerboards and
 ZTPG are observed by scanning tunnelling microscopy 
or spectroscopy, which can 
mainly see LDOS of the topmost  CuO$_2$ layer on a cleaved surface.
A possible explantation is that AF moments appear only in  few surface CuO$_2$ layers
because of surface defects or disorder.
It is likely that disorder enhances magnetism.
Experimentally, for example,
the doping of Zn ions enhances magnetism
in the vicinity of Zn ions \cite{Alloul};
theoretically,  magnetism is also enhanced by  disorder \cite{disorder}.

It is reasonable that checkerboards appear around
vortex cores because AF moments are induced there
\cite{vortex}; vortex cores can play a role of impurities for
quasiparticles as doped Zn ions do.
It is also reasonable that 
almost homogeneous checkerboards appear
in under doped cuprates 
where superconductivity disappears \cite{Hanaguri,Momono}; 
almost homogeneously AF moments  presumably exist there.
Inhomogeneous checkerboards can appear in under doped cuprates with rather low
SC $T_c$'s, 
when inhomogeneous AF moments are induced by disorder.

Even if  the $2{\bf Q}$ modulation in LDOS,
$\rho_{2{\bf Q}} (\varepsilon)$, is large,
the $2{\bf Q}$ electron density modulation or the amplitude of
CDW  caused by it is small
because of the cancellation between positive and negative parts of 
$\rho_{2{\bf Q}} (\varepsilon)$ below the chemical potential
and the small spectral weight of Gutzwiller's quasiparticles,
which is as small as  $|\delta|$,  with $\delta$ 
doping concentrations measured from the half filling,
according to Gutzwiller's theory \cite{Gutzwiller}.
On the other hand,    CDW 
can be induced by another mechanism in Kondo lattices \cite{multi-Q}.
In the vicinity of the Mott-Hubbard transition, 
local quantum spin fluctuations mainly quench
magnetic moments; this quenching is nothing but the Kondo effect.
The energy scale of local quantum spin fluctuations,
which is called the Kondo temperature $k_BT_K$,
depends on the electron filling in such a way that it is smaller 
when the filling is closer to the half filling; $k_BT_K \simeq |\delta t|$
according to Gutzwiller's theory \cite{Gutzwiller}.
Then, CDW is induced in such a way that
 $k_BT_K$ are smaller or local electron densities are closer to the half filling where
AF moments are larger. Even if the amplitude of CDW  
is not vanishingly small in an observation, 
the observation cannot contradict the mechanism of 
checkerboards and ZTPG proposed in this Letter.

In conclusion, double-{\bf Q} SDW with
${\bf Q}_1 = (\pm\pi/a,$  $\pm3\pi/4a)$ and 
${\bf Q}_2 = (\pm3\pi/4a,$ $\pm\pi/a)$ must be responsible
for  the so called $4a\times4a$
checkerboards in cuprate oxide superconductors,
with $a$ the lattice constant of CuO$_2$ planes.
Not only Cooper pairs with zero total momenta but also those with 
 $\pm2{\bf Q}_1$, 
$\pm2{\bf Q}_2$,  $\pm4{\bf Q}_1$, $\pm4{\bf Q}_2$, and so on
are possible in the SDW phase.
The so called zero temperature pseudogap phase must be
a coexisting phase of the double-${\bf Q}$ SDW and
the multi-{\bf Q} condensation of 
$d\gamma$-wave Cooper pairs.

The author is thankful for discussion to M. Ido, M. Oda, and N. Momono.


\begin{thebibliography}{}
\bibitem{bednorz}
J.G. Bednortz and K.A. M\"{u}ller, 
Z. Phys. B {\bf 64}, 189 (1986).
%
\bibitem{spingap}
H. Yasuoka et al.,
{\it Strong Correlation and Superconductivity},
Springer Series in Solid State Science Vol. 89
(Springer-Verlag, Berlin, New York 1989), p. 254.
%
\bibitem{Ding}
H. Ding et al.,
Nature {\bf 382}, 51 (1996).  
%
%
\bibitem{Shen2} 
J. M. Harris et al.,
Phys. Rev. B {\bf 54}, R15665 (1996).   
%
\bibitem{Shen3} 
A. G. Loeser,  et al.,
Science {\bf 273}, 325 (1996).  
%
\bibitem{Ino}
A. Ino et al.,
Phys. Rev. B {\bf 65}, 094504 (2002).
%
\bibitem{Renner}
Ch. Renner et al.,
Phys. Rev. Lett. {\bf 80}, 149 (1998). 
%
\bibitem{Ido1}
M. Oda et al.,
Physica C {\bf 281}, 135 (1997).
%
\bibitem{Ido2}
T. Nakano et al.,
J. Phys. Soc. Jpn. {\bf 67}, 2622 (1998).
%
%
%
\bibitem{Ekino}
T. Ekino et al.,
Phys. Rev. B {\bf 60}, 6916 (1999).
%
%
\bibitem{vortex}
J.E. Hoffman et al.,
Science {\bf 295}, 466 (2002).
%
\bibitem{Vershinin} 
M. Vershinin et al.,
Science, {\bf 303}, 1995 (2004).
%
\bibitem{Hanaguri}
T. Hanaguri et al.,
Nature {\bf 430}, 1001 (2004).
%
\bibitem{Howald}
C. Howald et al.,
Phys. Rev. B {\bf 67}, 014533 (2003).
%
\bibitem{Momono}
N. Momono et al.,
J. Phys. Soc. Jpn. {\bf 74}, 2400 (2005).
%
\bibitem{McElroy}
K. McElroy et al.,
Phys. Rev. Lett., {\bf 94}, 197005 (2005).
%
%
%
%
%
\bibitem{Mermin}
N.D. Mermin and H. Wagner,
Phys. Rev. Lett. {\bf 17}, 1133 (1966).
%
\bibitem{PS-gap}
F.J. Ohkawa,
cond-mat/0508344.
%
\bibitem{Davis}
H.C. Fu et al.,
cond-mat/0403001.
%
\bibitem{Sachdev}
S. Sachdev,
Science {\bf 288}, 475 (2000).
%
\bibitem{Vojta}
M. Vojta,
Phys. Rev. B {\bf 66}, 104505 (2002).
%
\bibitem{halperin}
D. Podolsky et al.,
Phys. Rev. B{\bf 67}, 094514 (2003).
%
\bibitem{cdw}
H.-D. Chen et al.,
Phys. Rev. Lett. {\bf 93}, 187002 (2004).
%
\bibitem{Chen1}
H.-D. Chen et al., 
cond-mat/040232.
%
\bibitem{Chen2}
H.-D. Chen et al., 
Phys. Rev. Lett. {\bf 89}, 137004 (2002).
%
\bibitem{Tesanovic}
Z. Tesanovi\'{c},  cond-mat/0405235.
%
\bibitem{Anderson}
P. W. Anderson,
cond-mat/0406038.
%
\bibitem{slave}
F.J. Ohkawa,
J. Phys. Soc. Jpn. {\bf 58}, 4156 (1989).
%
\bibitem{Hubbard} 
J. Hubbard, 
Proc. Roy. Soc. London Ser. A {\bf 276}, 238 (1963); A {\bf 281}, 401 (1964).
%
\bibitem{Gutzwiller} 
M.C. Gutzwiller, 
Phys. Rev. Lett. {\bf 10}, 159 (1963); Phys. Rev. A {\bf 134}, 293 (1963); 
 A {\bf137}, 1726 (1965).
%
\bibitem{Mapping-1} 
F.J. Ohkawa, 
Phys. Rev. B {\bf 44}, 6812 (1991).
%
\bibitem{Mapping-2} 
F.J. Ohkawa, 
J. Phys. Soc. Jpn. {\bf 60}, 3218 (1991); {\bf 61}, 1615 (1992).
%
\bibitem{georges}
A. Georges and G. Kotliar,
Phys. Rev. B {\bf 45}, 6479 (1992).
%
\bibitem{disorder}
F.J. Ohkawa, 
cond-mat/0510377.
To be published in J. Phys. Soc. Jpn.
%
\bibitem{hirsch}
J.E. Hirsch,
Phys. Rev. Lett. {\bf 54}, 1317 (1985).
%
%
\bibitem{KL-theory1} 
F.J. Ohkawa,
Jpn. J. Appl. Phys. {\bf 26}, L652 (1987);
J. Phys. Soc. Jpn. {\bf 56}, 2267 (1987).
%
\bibitem{KL-theory2} 
F.J. Ohkawa,
Phys. Rev. B. {\bf 59},  8930 (1990).
%
%
%
%
\bibitem{com}
A superconducting state with $\Delta_{l}\ne 0$ for $l\ne 0$
is nothing but a pair-density wave state  or a generalized
Fulde-Ferrel-Larkin-Ovchinnikov (FFLO) state induced by AF fields.
Electron pairs of ${\bf k}+ (l+\frac1{2}) {\bf Q}$ and
$-{\bf k}+(l+\frac1{2}){\bf Q}$, whose total momenta are
$(2l+1){\bf Q}$, are also possible. However,
they are ignored here; 
they should be considered in case of CDW.
%
\bibitem{ZhangRice}
F.C. Zhang and T.M. Rice,
Phys. Rev. B {\bf 37}, R3759 (1988). 
%
%
\bibitem{gamma1} 
J.W. Loram, et al.,
Phys. Rev. Lett. {\bf 71}, 1740 (1993).
%
\bibitem{multi-Q}
F.J. Ohkawa, J. Phys. Soc. Jpn. {\bf 67}, 535 (1998);
Phys. Rev. B {\bf 66}, 014408 (2002).
%
\bibitem{Alloul}  
H. Alloul et al., 
Phys. Rev. Lett. {\bf 67} (1991) 3140.
%
%
%
%
\end{thebibliography}
\end{document}